\documentstyle
[12pt]{article} \hoffset -0.5in \textwidth 6.5in \textheight
9.00in  \parskip 7pt \openup2.5\jot \parindent=0.5in
\renewcommand{\thefootnote}{\fnsymbol{footnote}}

\def\lsim{\;\centeron{\raise.35ex\hbox{$<$}}{\lower.65ex\hbox
{$\sim$}}\;}
\def\gsim{\;\centeron{\raise.35ex\hbox{$>$}}{\lower.65ex\hbox
{$\sim$}}\;}
\def\centeron#1#2{{\setbox0=\hbox{#1}\setbox1=\hbox{#2}\ifdim
\wd1>\wd0\kern.5\wd1\kern-.5\wd0\fi
\copy0\kern-.5\wd0\kern-.5\wd1\copy1\ifdim\wd0>\wd1
\kern.5\wd0\kern-.5\wd1\fi}}

\newskip\humongous \humongous=0pt plus 1000pt minus 1000pt

\newif\ifdtup

\relax
\def\np#1#2#3 {Nucl. Phys. \underline{#1} (19#2) #3}
\def\prl#1#2#3 {Phys. Rev. Lett. \underline{#1} (19#2) #3}
\def\prev#1#2#3 {Phys. Rev. \underline{#1} (19#2) #3}
\def\pl#1#2#3 {Phys. Lett. \underline{#1} (19#2) #3}
\def\rmp#1#2#3 {Rev. Mod. Phys. \underline{#1} (19#2) #3}
\def\prep#1#2#3 {Phys. Rep. \underline{#1} (19#2) #3}
\def\spu#1#2#3 {Sov. Phys.-Usp. \underline{#1} (19#2) #3}
\def\sjnp#1#2#3 {Sov. J. Nucl. Phys. \underline{#1} (19#2) #3}
\def\jetp#1#2#3 {JETP Lett. \underline{#1} (19#2) #3}
\def\apj#1#2#3 {Astrophys. J. \underline{#1} (19#2) #3}
\def\apjl#1#2#3 {Astrophys. J. Lett. \underline{#1} (19#2) #3}
\def\ib#1#2#3 {{\it ibid.} \underline{#1} (19#2) #3}
\def\nat#1#2#3 {Nature (London) \underline{#1} (19#2) #3}
\def\ap#1#2#3 {Ann. Phys. (NY) \underline{#1} (19#2) #3}
\def\nc#1#2#3 {Nuovo Cim. \underline{#1} (19#2) #3}
\def\zp#1#2#3 {Zeit. Phys. \underline{#1} (19#2) #3}
\def\ar#1#2#3 {Ann. Rev. Nucl. Part. Sci. \underline{#1} (19#2) #3}
\def\prs#1#2#3 {Proc. Roy. Soc. \underline{#1} (19#2) #3}
\def\pcps#1#2#3 {Proc. Cam. Phil. Soc. \underline{#1} (#2) #3}
\def\rpp#1#2#3 {Rep. Prog. Phys. \underline{#1} (19#2) #3}
\def\cpc#1#2#3 {Computer Phys. Comm. \underline{#1} (19#2) #3}
\def\ptp#1#2#3 {Prog. Th. Phys. \underline{#1} (19#2) #3}
\def\ijmp#1#2#3 {Int. J. Mod. Phys. \underline{#1} (19#2) #3}
\def\app#1#2#3 {Acta. Phys. Pol. \underline{#1} (19#2) #3}
\def\mpl#1#2#3 {Mod. Phys. Lett. \underline{#1} (19#2) #3}
\def\beq{\begin{equation}}
\def\eeq{\end{equation}}

\def\VEV#1{\left\langle #1\right\rangle}

\relax
\begin{document} \begin{titlepage}
\rightline{\vbox{\halign{&#\hfil\cr
&ANL-HEP-CP-93-56\cr
&\today\cr}}}
\vspace{0.25in}
\begin{center}

{\Large\bf
THE MASS OF THE HEAVY AXION $\eta_6$}
\medskip

Alan R. White\footnote{Work
supported by the U.S. Department of
Energy, Division of High Energy Physics, Contract\newline W-31-109-ENG-38}
\\ \smallskip
High Energy Physics Division\\Argonne National
Laboratory\\Argonne, IL 60439\\ \end{center}

\begin{abstract}
      If electroweak dynamical symmetry breaking is due to a chiral
condensate of color sextet quarks, dynamics analagous to ``walking
technicolor'' will enhance the condensate by orders of magnitude compared to
the electroweak chiral scale. This enhancement compensates for the
exponential suppression of electroweak scale color instanton interactions.
As a result the $\eta_6$ axion can naturally aquire an electroweak scale
mass.

\vspace{3.0in}
\noindent Contributed to the XVI International Symposium on Lepton-Photon
Interactions, ~~~~Cornell University, Aug. 10-15th, 1993.

\end{abstract}

\renewcommand{\thefootnote}{\arabic{footnote}} \end{titlepage}

The existence of a massive color sextet quark sector could be an essential
factor in the very high-energy consistency of QCD \cite{arw} and in the
dynamics of the electroweak sector \cite{sex,kkw}. In particular, low-energy
Strong CP conservation in the normal triplet quark sector can be directly
due to a sextet quark axion state\cite{kkw}, the $\eta_6$. If this is the
case the QCD interactions of the sextet sector are not CP conserving. As a
result the $\eta_6$ has large CP violating couplings to the electroweak
sector that could be responsible for its production at LEP as an
intermediate state in $Z^0 \to \gamma\gamma + \mu^+\mu^-$ events \cite{kkw}.
Mixing of the color triplet and sextet sectors could also underlie CP
violation in general.

At first sight\cite{ros}, the $\eta_6$ is a conventional Peccei-Quinn
axion\cite{pec}. However, because of its higher color constituents, color
instanton interactions provide an additional contribution to its mass
\cite{hol} not envisaged in the original Peccei-Quinn mechanism. It is very
important to understand how large this contribution can be. In this paper
we shall outline how the dynamics\cite{app} of ``walking technicolor''(WTC), as
applicable to sextet electroweak symmetry breaking, can compensate for the
normal suppression of instanton interactions. As a result it is natural to
expect the $\eta_6$ to have an electroweak scale mass.

We shall suppose that the entry of a flavor doublet (U,D) of color sextet
quarks into QCD above the electroweak scale can be described by an effective
$\beta$-function. If we write
\beq
\label{beta}
\beta(\alpha) = - \beta_0 \alpha^2(q)/2\pi~ -~ \beta_1 \alpha^3(q)/8\pi^2 +
{}~~...
\eeq
then for six color triplet flavors the normal two-loop calculation gives
\beq
\label{six}
\beta_0 = 11 - 2n_f/3~ = 7,~~~\beta_1 = 102 - 38n_f/3~= 26
\eeq
whereas when the two sextet flavors are included we obtain\cite{tar}
\beq
\label{sex1}
\beta_0 = 7 - 4T(R)n^6_f/3~ = 7 - 4(\frac{5}{2})2/3~ = 1/3,
\eeq
and
\beq
\label{sex2}
\beta_1 = ~26 - 20T(R)n^6_f - 4C_2(R)T(R)n^6_f~ =~26 - 100 -66\frac{2}{3}
{}~=~-140\frac{2}{3}
\eeq
where we have used $T(R) = 5/2$ and $C_2(R) = 10/3$ for sextet quarks.

The resulting $\beta$-function, $\beta^{(6)}$, for six flavors of light triplet
quarks (we ignore subtleties associated with a heavy top quark) is shown in
Fig.~1(a) and compared, in Fig.~1(b), with $\beta^{(6,2)}$, - the
$\beta$-function obtained with the sextet quarks added. Noting the greatly
expanded vertical scale in Fig.~1(b), it is clear that as soon as the sextet
sector enters the theory the evolution of $\alpha_s$ essentially comes to a
halt. If, as we assume, the theory nevertheless evolves smoothly, but very
slowly, into the small coupling asymptotically-free region then a natural
way to connect the evolution before and after the sextet sector enters is
via a $\beta$-function of the form shown in Fig.~2. Such an evolution
(presumably) provides an oversimplified picture of the physics involved but
will allow us to give an order of magnitude  discussion of quantities which
we hope is not too unrealistic. We assume, therefore, that above the
electroweak
scale the $\beta$-function can be taken to be a small (almost) momentum
independent constant, which we denote as $\beta_c$. This provides the
essential prerequisite\cite{app} for the application of WTC dynamics.

A major ingredient of WTC, in addition to the existence of a small $\beta_c$,
is the assumption\cite{app} that the linearised Dyson-Schwinger ``gap
equation'' for the quark dynamical mass $\Sigma(p)$, gives a semi-quantitive
description of the dynamics of chiral symmetry breaking. The gap equation has a
solution corresponding to spontaneous chiral symmetry breaking for $\alpha_s
\geq \alpha_c$, where $\alpha_c$ is determined by an equation of the form
\beq
\label{chir}
C_2(R) \alpha_c = \mbox{constant}
\eeq
$C_2(R)$ is the Casimir operator already referred to above and, since $C_2$ for
sextet quarks is 5/2 the corresponding triplet Casimir, (\ref{chir}) is
consistent with the sextet chiral scale being the electroweak scale provided
that $\alpha_s$ evolves logarithmically from the triplet chiral scale up to
this scale, in the usual manner.

If $\alpha_s$ is momentum independent above the electroweak scale, as will
be approximately the case if $\beta_c$ is small enough, then the solution of
the gap equation for $\alpha_s \sim \alpha_c$ has the form\cite{app}
\beq
\label{sig}
\Sigma (p) \sim ~\mu^2~(p)^{-1}
\eeq
where $\mu$ is determined at the electroweak scale and should be
essentially the sextet quark constituent quark mass i.e. $\mu~\geq~ 300$ GeV.
When this behavior is inserted into the perturbative formula\cite{app} for the
high-momentum component of the sextet condensate $\VEV{Q\bar{Q}}$ we obtain a
contribution
\beq
\label{cond}
\VEV{Q\bar{Q}} \sim \int dp~p~ \Sigma(p) ~\sim \mu^2\Lambda
\eeq
where $\Lambda$ is the upper cut-off on the integral. In contrast, the
corresponding perturbative formula\cite{app,pag} for the chiral constant
$F_{\pi_6}$ (which determines the $W$ and $Z$ masses) gives a
high-momentum component
\beq
\label{ps}
F^2_{\pi_6} \sim \int dp~p^{-1}~\Sigma^2(p)~~+~~...
\eeq
which is not enhanced by the behavior (\ref{sig}).

The upper cut-off $\Lambda$ should naturally be of the same order as the scale
$M$ at which new (unification) physics appears. However, it could be greater
if the new physics is asymptotically-free and does not significantly change
the evolution of $\alpha_s$. The new physics must provide four-fermion
triplet/sextet couplings in particular, which then provide triplet quark
masses with the order of magnitude
\beq
\label{mas}
m_q \sim ~\VEV{Q\bar{Q}}/M^2 ~\sim ~\mu^2\Lambda/M^2
\eeq
To be sure that flavor-changing neutral currents are suppressed we must
certainly take\cite{app} ~$M \geq 300$ TeV. However, if $M = 300$ TeV,
then from (\ref{mas}) we see that to obtain a mass $\sim$ 10-100 Mev for a
triplet quark, we should have
\beq
\label{order}
\VEV{Q\bar{Q}} \sim~ 10^{10} - 10^{11}~(GeV)^3
\eeq
If we take $\mu^2 \sim~10^5~(GeV)^2$, we must then have
\beq
\label{lam}
\Lambda \sim~10^5 - 10^6~GeV~= 100 - 1,000~TeV.
\eeq
so that $\Lambda \sim M$ is clearly possible. Equations (\ref{chir}) -
(\ref{lam}) simply illustrate the essence of WTC, i.e. if the gauge
coupling starts to ``walk'' immediately above the chiral scale, then the
chiral condensate can be enhanced by orders of magnitude relative to the
chiral scale. This allows reasonable fermion masses to be generated without
accompanying problems from flavor-changing processes. As we now describe,
the instanton interactions contributing to the $\eta_6$ mass are also enhanced.

The $\eta_6$ is a Goldstone boson associated with the axial U(1) symmetry
orthogonal to that broken by color instantons. As such it is an axion which
remains massless until four-fermion $q\bar{q}Q\bar{Q}$ couplings are
added to the theory. Such couplings then combine with the triplet/sextet
instanton interactions and the sextet condensate to produce an $\eta_6$
mass. The one instanton contribution to this mass is illustrated in Fig.~3.

Because $T(r)$ is 5/2 for sextet quarks and 1/2 for triplet quarks, the
axial U(1) current that is conserved before the addition of four-fermion
couplings is
\beq
\label{cur}
J^A = 5J^{3A} - J^{6A}
\eeq
where $J^{3A}$ and $J^{6A}$ are respectively the triplet and sextet
U(1) currents. As a result the one instanton interaction appearing in Fig.~3
involves {\it five} sextet quarks (and antiquarks) of each flavor and just
one triplet quark (and antiquark) of each flavor. It is therefore a very
high-order fermion interaction of the form
\beq
\label{inst}
(U\bar{U})^5(D\bar{D})^5u\bar{u}d\bar{d}s\bar{s}c\bar{c}b\bar{b}t\bar{t}
\eeq
This interaction is scaled by a factor $(P_I)^{-44}$, where $P_I$ is the
momentum scale corresponding to the size of the instanton, and a
non-perturbative exponential suppression factor involving the instanton
action. Since instanton interactions above the electroweak scale are
well-defined (because of the absence of renormalons) we take the
minimum allowed value for $P_I$ to be the electroweak scale i.e.
$P_I~\sim~100$ GeV. (To be really sure that we do not induce interactions
violating the established success of perturbative $QCD$ we should probably
take $P_I$ slightly larger. However, our estimates wil be sufficiently crude
that this will not be significant. Also we shall overestimate other
suppression factors, for example by underestimating $\alpha_s$.)

Given our assumption that the scale of the instanton interaction is the
electroweak scale, we will take $\alpha_s \sim 10^{-1}$. In this case
the non-perturbative factor for the one instanton interaction is
\beq
\label{sup}
[2\pi/\alpha_s]^6~\exp{[-2\pi/\alpha_s]}~~ \sim~ [60]^6~\exp{[-60]}
{}~~\sim (10)^{-15}
\eeq
which, of course, provides a large suppression of the interaction.

It is crucial that the four-fermion couplings and sextet condensate attached
to the triplet quark legs of the instanton interaction combine to produce a
mass factor $m_q$ for each triplet quark (together with a factor of $P^2_I$
for each of the loop integrals involved). The result is an overall phase
factor of $\bar{\theta} = \theta + arg~det~m^{(3)}$, where $\theta$ is the
usual
topological parameter and $m^{(3)}$ is the triplet quark mass matrix. This
allows\cite{hol} a mass to be generated which does not disturb the CP
conserving minimum of the axion action at $\bar{\theta} = 0$. However, the
factor of $det~m^{(3)}$ (scaled by $(P_I)^6$) also provides a significant
suppression of the interaction. If we take, in order of magnitude,
$m_u ~\sim~m_d~\sim~10^{-2}$ GeV, $m_s ~\sim~10^{-1}$ GeV, $m_c ~\sim~1$ GeV,
$m_b ~\sim~10$ GeV, and $m_t ~\sim~100$ GeV,
we can estimate this suppression as
\beq
\label{det}
(det~m_3)~/ (100~GeV)^6~~
\sim~(10^{-2})^2(10^{-1})(10)(10^2)~/~(10^2)^6
{}~\sim(10)^{-14}
\eeq
which is a comparable suppression to (\ref{sup}). Both (\ref{sup}) and
(\ref{det}) are automatically present in electroweak scale instanton
interactions and provide the usual reason for regarding such interactions as
completely negligible.

Now we come to the role of the sextet condensate in compensating for the
suppression factors (\ref{sup}) and (\ref{det}). If we use the
sextet dynamical mass $\Sigma(p)$ to tie together sextet quark legs of the
instanton interaction, then the propagators present simply give the high
momentum integral (\ref{cond}) and so produce the condensate as an overall
multiplicative factor, as illustrated in Fig.~3. As a result, if the condensate
enhancement is as large as (\ref{order}), the combined suppression of
(\ref{sup}) and (\ref{det}) is overcome. The higher color of the sextet
quarks is {\it vital} here since it is responsible for the large power of the
condensate in the single instanton interaction. Indeed, in the $\eta_6$
mass, we have an an enhancement factor of $\VEV{Q\bar{Q}}^8$ which, using
(\ref{order}), and including the remaining factor of $(P_I)^{-26}$, gives
\beq
\label{enha}
(\VEV{Q\bar{Q}})^8/~(100~GeV)^{26}~~~\sim~  (10^{10} - 10^{11})^8/~(10)^{52}
{}~\sim~(10)^{28} - (10)^{36}
\eeq
and clearly this can be more than sufficient to compensate for the
combined suppression due to (\ref{sup}) and (\ref{det}). Of course, our
estimates are very crude and the numbers could vary considerably from those
we have used. However, that the condensate enhancement can reasonably be
expected to enhance the instanton interaction from a very weak to a strong
interaction is surely well illustrated by the above discussion.

If the condensate is large enough to produce a strong one instanton
interaction then clearly multi-instanton interactions will also be strong
and we will have a highly non-perturbative theory. Indeed since renormalons
are probably absent\cite{arw} in the full triplet plus sextet version of QCD,
instantons provide the only non-perturbative aspect of the theory. It could
be, therefore, that the large sextet condensate is necessary to enhance these
interactions sufficiently to produce a ``non-perturbative'', confining,
solution of the theory at the electroweak scale which is able to match with
a low-energy confined triplet sector.

That the instanton interactions are enhanced presumably invalidates the
starting assumption of WTC that the dynamics of chiral symmetry breaking is
well-described by the linearised Dyson-Schwinger equation. However, from our
point of view, this assumption is simply a way of selecting a set
of interactions that can produce the dynamics we are proposing. We
acknowledge that a full description of the dynamics is surely much more
complicated than we have described but, as we stated above, we hope that our
order of magnitude discussion retains some substance.

If the instanton interactions provide the major dynamics of the theory then
the $\eta_6$ mass should automatically inherit the main dynamical scale of
the theory. In this case an electroweak scale mass of, say, 60 GeV would be
quite natural. We should also note that, in principle at least, the overall
magnitude of the four-fermion triplet/sextet couplings can be regarded as a
parameter which can be smoothly varied from zero to its physical value.
During this variation the $\eta_6$ mass will go from zero to its physical
value. Since the axion status of the $\eta_6$ will be preserved throughout
the variation, it is clear that Strong CP will be strictly conserved by the
triplet quark sector as masses are induced (by the four-fermion couplings)
and that this will continue to be the case even if the $\eta_6$ mass is very
large.

It seems therefore that the $\eta_6$ is a good candidate for the new massive
state that might have been observed\cite{kkw} at LEP - via its two photon
decay mode in particular. It is important to emphasize in this context that
the underlying dynamics of the sextet sector, which couples the $\eta_6$ to
the $W^+$, $W^-$ and $Z^0$ is novel in that it involves enhanced instanton
interactions. This is why the operators involved in the decay modes are,
perhaps, unexpected. Indeed papers have been written, for example
\cite{lis}, arguing that a new scalar particle interpretation of the LEP
events is implausible because the operators that could be involved would
also appear in other experiments - where they can clearly be ruled out. In
fact there is no inconsistency between the candidate LEP events and other
experiments if the events are regarded as $\eta_6$ production. In the
classification (and words) of \cite{lis}, the operator involved is of the
``$O_2$ - type'' which ``only contain $Z$ and $W$ gauge bosons and ... this
fact prevents us from ruling them out with current experimental data''.

\newpage
\noindent{\bf Figure Captions}

\begin{itemize}

\item[{Fig.~1}] (a) The $\beta$-function $\beta^{(6)}$, for $QCD$ with six
light flavors of color triplet quarks, and (b) comparison of $\beta^{(6,2)}$,
the $\beta$-function with two color sextet flavors added, with
$\beta^{(6)}$.

\item[{Fig.~2}] The effective $\beta$-function describing the evolution as
the sextet sector enters the theory.

\item[{Fig.~3}] The one instanton interaction contributing to the $\eta_6$
mass.

\end{itemize}

\end{document}